\documentclass{svproc}
\usepackage{url}
\usepackage{cite}
\usepackage{amsmath,amssymb,amsfonts}
\usepackage{algorithmic}
\usepackage{textcomp}
\usepackage{xcolor}
\usepackage{graphicx}
\usepackage{threeparttable}
\usepackage{array}
\usepackage{colortbl}

\usepackage{amssymb}
\usepackage{pifont}
\newcommand{\cmark}{\ding{51}}%
\usepackage{enumitem}
\usepackage[bookmarks=false]{hyperref}
\newcommand*\xor{\mathbin{\oplus}}

\begin{document}
\mainmatter              
\title{abstractPIM: A Technology Backward-Compatible Compilation Flow for Processing-In-Memory}
\titlerunning{A Technology Backward-Compatible Compilation Flow for PIM}  
%
\author{Adi Eliahu \and Rotem Ben-Hur
\and Ronny Ronen \and Shahar Kvatinsky}
\authorrunning{Adi Eliahu et al.} 
%
\tocauthor{Adi Eliahu, Rotem Ben-Hur, Ronny Ronen, and Shahar Kvatinsky}
\institute{\textit{Technion - Israel Institute of Technology}\\
Haifa, Israel 3200003,\\
\email{\{adieliahu, rotembenhur\}@campus.technion.ac.il, ronny.ronen@technion.ac.il, shahar@ee.technion.ac.il},\\ 
}

\maketitle              

\begin{abstract}
The von Neumann architecture, in which the memory and the computation units are separated, demands massive data traffic between the memory and the CPU. 
To reduce data movement, new technologies and computer architectures have been explored. The use of memristors, which are devices with both memory and computation capabilities, has been considered for different processing-in-memory (PIM) solutions, including using memristive stateful logic for a programmable digital PIM system.
Nevertheless, all previous work has focused on a specific stateful logic family, and on optimizing the execution for a certain target machine. These solutions require new compiler and compilation when changing the target machine, and provide no backward compatibility with other target machines.
In this chapter, we present abstractPIM, a new compilation concept and flow which enables executing any function within the memory, using different stateful logic families and different instruction set architectures (ISAs). By separating the code generation into two independent components, intermediate representation of the code using target independent ISA and then microcode generation for a specific target machine, we provide a flexible flow with backward compatibility and lay foundations for a PIM compiler.
Using abstractPIM, we explore various logic technologies and ISAs and how they impact each other, and discuss the challenges associated with it, such as the increase in execution time. 

\keywords{Memristor, processing-in-memory, RRAM, stateful logic, ISA}
\end{abstract}
\section{Introduction}
In recent years, the trend of data-intensive applications has become popular. Data-intensive applications process large volumes of data and also exhibit compute-intensive properties, and therefore, they require massive data transfer between the memory and the central processing unit (CPU). Since there is a large performance gap between the CPU and the memory~\cite{Pedram2017}, this massive data transfer has become a bottleneck in execution of data-intensive applications. This bottleneck is often called the \textit{memory wall}. As a result of the memory wall challenge, processing-in-memory (PIM) has become attractive~\cite{MemristorMythOrReality,InMemoryCompReview}.
Due to the recent advances in memory technologies, \textit{e.g.}, resistive random access memory (RRAM)~\cite{RRAM} and PCM~\cite{PCM}, PIM has gained interest and has become an integral part of many computer architectures.
The memristor, which functions as both a memory element and a computation unit, can help reducing data transfer between the CPU and the memory and thus addresses the memory wall problem.
By applying voltage across the device, the memristor performs switching between two resistance values, high resistance value ($R_{OFF}$) and low resistance value ($R_{ON}$), therefore it can function as a binary memory element. 
To increase the memristor density, it can be programmed to have intermediate resistance between $R_{OFF}$ and $R_{ON}$, thus achieving multi-level cell (MLC) storage capability.

In addition to their storage capabilities, memristors can also be used for computation. There are two approaches to use memristors as computation units. The first approach is using the memristor in application-specific architectures. Memristors can be used for the purpose of a specific computation. For example, in~\cite{MatrixVectorMult}, an efficient vector-matrix multiplication using memristor analog computation is demonstrated.
In this manner, the dual-function memristor can perform efficient computing and reduce data transfer requirements between the CPU and the memory.
Numerous accelerators integrating analog memristor-based computations have recently been developed, mostly for artificial intelligence applications~\cite{acceleratorSurvey}. 

The second approach of using memristor as a computation unit, on which we focus in this chapter, is called 'stateful logic'. Using stateful logic, memristive logic gates are constructed within the memory array for general-purpose computation. Stateful logic enables programmable general-purpose architectures since every memristive cell can be used as a storage element, as well as an input, output or a register. Several memristor logic gate families have been designed, including MAGIC~\cite{MAGIC}, IMPLY~\cite{IMPLYNature}, and resistive majority~\cite{MIG}.

Some stateful logic families can be easily integrated within a memristive crossbar array with minor modifications. Designing a functionally complete logic gate set using such a family, \textit{e.g.}, a MAGIC NOR gate, enables in-memory execution of any function. 
Various logic gate families have been explored in the literature, each of them has different advantages. Previous efforts to execute a function within the memory concentrated on utilizing a specific PIM family and optimizing the latency, area, or throughput using this technology, \textit{e.g.}, SAID~\cite{SAID} and SIMPLE~\cite{SIMPLE} for optimizing latency in MAGIC technology~\cite{MAGIC}, SIMPLER~\cite{SIMPLER} for optimizing throughput in MAGIC technology~\cite{MAGIC} and K-map based synthesis~\cite{IMPLYSynthesis} for optimizing latency and area in IMPLY~\cite{IMPLYNature}. 

While these previous works have considerably improved the logic function execution in terms of latency, area, or throughput, they are strongly dependent on the PIM family and its basic operations, and therefore are limited to a specific target machine. However, each PIM technology has different advantages, and therefore, flexibility in the used PIM technology has many motivations. For example, the MAGIC family provides memristive crossbar compatibility and high parallelism by executing MAGIC logic gates on aligned elements in different rows of the memristive crossbar. A different PIM technology, called CRS~\cite{CRS}, provides flexibility by executing 16 Boolean functions in a single operation.

In this chapter, we propose a new hierarchical compilation method for PIM, which provides flexibility and is not restricted to a certain PIM technology. Our flow separates the code generation into two components. The first component is intermediate code generation using target independent instruction set architecture (ISA). The second component is microcode generation for a specific target machine and PIM technology. The third component, runtime execution, executes the code. The first component, which is run by the programmer, is independent of the PIM technology. In this component, a compiled program that consists of target-independent instructions is generated. 
In the second component, which is target-dependent, these instructions are translated into an execution sequence of micro-operations supported by the target machine. The second component is performed by the PIM technology provider. In the third component, at runtime, the compiled code instructions are sent from the CPU to the memory controller, which contains the instruction execution sequences from the second component. The controller translates the instructions into micro-operations and sends them to the memory. This third component is similar to an instruction-level opcode being executed using micro-operations in the x86 processors~\cite{ProcessorMicroarchitecture2}.

\begin{figure}[!t]
    \centering
    \includegraphics[width=1\columnwidth,clip]{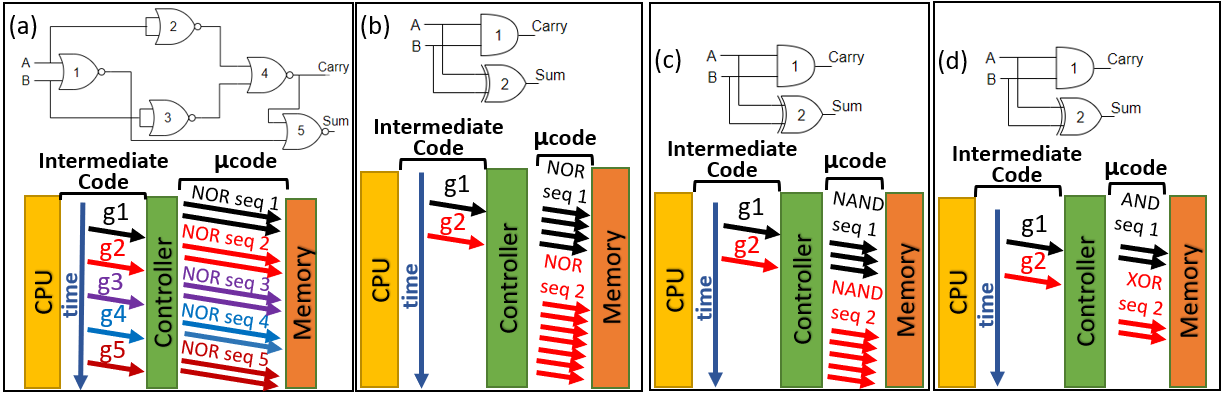}
    \caption{Compilation example for a half adder using various ISAs and target machines. (a) A NOR ISA and MAGIC NOR target machine. (b) All 2-input and single-output ISA and MAGIC NOR target machine. (c) All 2-input and single-output ISA and MAGIC NAND target machine. (d) All 2-input and single-output ISA and 2-input and single-output MAGIC target machine.}
    \label{fig:controlLoad}
\end{figure}

Figure~\ref{fig:controlLoad} demonstrates the first and third flow components of a half adder logic for different ISAs and target machines. The first two implementations, shown in Figure~\ref{fig:controlLoad}(a) and~\ref{fig:controlLoad}(b), demonstrate the use of the same target machine while using different ISAs. The code is compiled for a machine that its PIM technology supports only MAGIC NOR logic gates. However, the first example targets a controller which supports only NOR ISA commands, whereas the second example supports all the 2-input and 1-output logic functions as its ISA. In the first component, a netlist and compiled program composed of the ISA commands, dubbed \textit{instructions}, are generated. In Figure~\ref{fig:controlLoad}(a), the netlist is composed of five logic gates that implement the half adder logic, and in Figure~\ref{fig:controlLoad}(b) it is composed of two gates (AND and XOR). The number of gates in the netlist is a representative of both the code size (or number of commands sent from the CPU to the PIM machine), and the control load between the CPU and the memory controller. We will refer to it for the rest of the chapter as \textit{code size}. The code size is also a means of estimation of the code abstraction achieved by our flow. In these examples, the code sizes are five and two, respectively. The second component is the microcode generation, where each command is translated to a sequence of MAGIC NOR operations and is embedded in the controller. In the third component, the code is executed. The commands are sent from the CPU to the controller, and then from the controller to the memory; hence, the code size is reduced with minimal changes to the in-memory implementation, namely, adding a few states to the memory controller to support other operations. 

Figures~\ref{fig:controlLoad}(b), ~\ref{fig:controlLoad}(c) and~\ref{fig:controlLoad}(d) demonstrate the use of the same ISA while using different target machines. These three examples use all 2-input logic functions as their ISA, but the first machine uses MAGIC NOR technology, the second uses MAGIC NAND technology and the third uses all MAGIC 2-input logic functions. This example demonstrates the ISA definition flexibility and command hierarchy enabled by our method, and the possible reduction in code size and reduction in the control load between the CPU and the memory controller. It also demonstrates the backward compatibility feature; in Figures~\ref{fig:controlLoad}(c)-(d), machines with technologies which enable lower execution time are used, and yet the generated intermediate code is backward compatible with other PIM technologies. The separation into two independent code generation components also enables the exploration of the impact of the ISA on the used target machine and vice versa.

This chapter makes the following contributions:
\begin{enumerate}[leftmargin=.15in]
\item Development of technology-independent and ISA-flexible flow (first presented in~\cite{abstractPIM}) for executing any logic function to a memristive crossbar array. Our technique, called abstractPIM, presents a hierarchical view and includes three components. It is a solid foundation for implementation of compilers for general-purpose memristive PIM architectures. This chapter also extends the work in~\cite{abstractPIM} and discusses future work of the abstractPIM flow.
\item Examining the impact of the ISA and the target machine on each other using abstractPIM, in terms of flexibility, performance and code size.
\item A 56\% reduction in the control load between the CPU and the memory controller as compared to state-of-the-art solutions~\cite{SIMPLER}, demonstrated for different benchmarks.
\end{enumerate}
\section{Background and Related Work}
\subsection{Stateful Logic}
\begin{figure}[!t]
    \setlength{\abovecaptionskip}{0pt}
    \centering
    \includegraphics[width=0.7\columnwidth,clip]{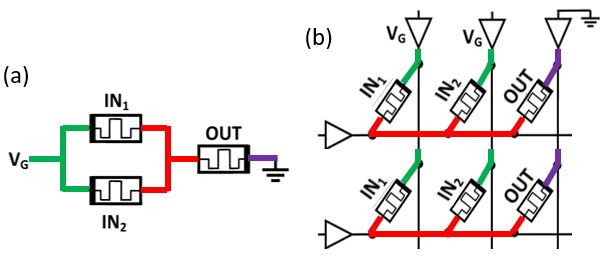}
    \caption{MAGIC NOR gates. (a) MAGIC NOR gate schematic. (b) Two MAGIC NOR gates in a crossbar configuration, executed in parallel.
    }
    \label{fig:MAGIC}
\end{figure}

In stateful logic families~\cite{JohnPatmos}, the logic gate inputs and outputs are represented by memristor resistance. We demonstrate the stateful logic operation using MAGIC~\cite{MAGIC} gates, which are used as a baseline in this chapter. Figure~\ref{fig:MAGIC}(a) depicts a MAGIC NOR logic gate; the gate inputs and output are represented as memristor resistance. The two input memristors are connected to an operating voltage, $V_{g}$, and the output memristor is connected to the ground. The output memristor is initialized at $R_{ON}$ and the input memristors are set with the input values. During the execution, the resistance of the output memristor changes according to the ratio between the input values and the initialized value at the output. For example, when one or two inputs of the gate are logical '1', according to the voltage divider rule, the voltage across the output memristor is higher than \(\frac{V_{g}}{2}\). This causes the output memristor to switch from $R_{ON}$ to $R_{OFF}$, matching the NOR function truth table. The MAGIC NOR gate can be integrated in a memristive crossbar array row, as shown in Figure \ref{fig:MAGIC}(b). Integration within the crossbar array enables executing logic gates in different rows in the same clock cycle, thus providing massive parallelism.

\subsection{Logic Execution within a Memristive Crossbar Array}
\label{subsec:logic_execution}
In CMOS logic, execution of an arbitrary logic function is performed by signals propagating from the inputs towards the outputs. However, in stateful logic, the execution is performed by a sequence of operations, each operation operates in a single clock cycle. In each clock cycle, one operation can be performed either on a single row, or on multiple rows concurrently. Overall, the execution takes several clock cycles. A valid logic execution is defined by mapping of every logic gate in the desired function to several cells in the crossbar array, and operating it in a specific clock cycle. 

\begin{figure}[!t]
    \setlength{\abovecaptionskip}{0pt}
    \centering
    \includegraphics[width=1\columnwidth,clip]{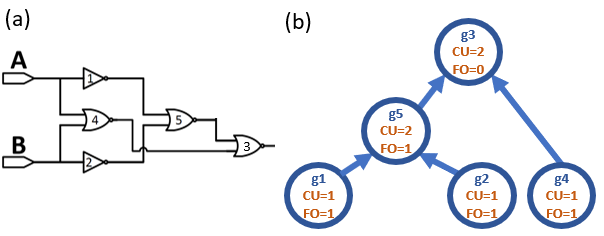}
    \caption{SIMPLER \textit{CU} and \textit{FO} node values example. (a) An example netlist. (b) The SIMPLER DAG generated from the example netlist, including the \textit{CU} and \textit{FO} values.
    }
    \label{fig:SIMPLER_CU}
\end{figure}

Many tools to generate the sequence of operations and map them into the memristive crossbar array cells have been discussed in the literature, \textit{e.g.}, SIMPLE~\cite{SIMPLE}, SAID~\cite{SAID}, the tool suggested by Yadav \textit{et al.}~\cite{YADAV} and the tool suggested by Thangkhiew \textit{et al.}~\cite{BDDMapping}. These tools map the logic to several rows in the memristive crossbar. Recently, a new method, called SIMPLER~\cite{SIMPLER}, which maps the logic to a single row, has been presented. This method tries to minimize the number of initialization cycles in the execution sequence to reduce the overall number of execution cycles. The input logic function of SIMPLER is synthesized using the ABC synthesis tool \cite{ABC}, which generates a netlist implementing the function with NOT and NOR gates only. Then, an in-house mapping tool builds a directed acyclic graph (DAG), in which every node represents a gate in the netlist. Each node is given two values: fanout (\textit{FO}) and cell usage (\textit{CU}). The former indicates how many parents the node has (\textit{i.e.}, how many gates are directly connected to its output), and the latter estimates the cell usage of the sub-graph starting from the node (\textit{i.e.}, the number of cells necessary for the execution of the sub-graph). An example of the \textit{CU} and \textit{FO} node values is demonstrated in Figure~\ref{fig:SIMPLER_CU}. Figure~\ref{fig:SIMPLER_CU}(a) shows a netlist, and Figure~\ref{fig:SIMPLER_CU}(b) shows its corresponding SIMPLER DAG with the \textit{CU} and \textit{FO} node values. 

The $CU$ of a node $V$ is calculated by:
\begin{itemize}
    \item If $V$ is a leaf then: 
    \begin{equation} CU(V)=1 \end{equation}
    \item Else, sort all $N$ children of $V$ by descending order of their $CU$ values. Then:
    \begin{equation}CU(V)=max\{CU(V_{child,i}))+i-1\},\forall i=(1\ to\ N)\end{equation}
\end{itemize}

Based on the \textit{CU} and \textit{FO} values, the mapping tool determines the order in which the gates operate. Additionally, the mapping tool traces the number of available cells, and when they are all occupied, it adds an initialization cycle in which cells are initialized and then reused. The gate execution ordering is determined such that the number of initialization cycles, and consequently overall execution time, will be minimized.

The gap between target machine constraints and architectural design choices, \textit{e.g.}, ISA, has not been addressed in the aforementioned mapping tools. Attempts have been made in existing mapping tools to support complex operations in the in-memory execution, \textit{e.g.}, 4-input LUT function~\cite{SAID}. However, their flexibility is limited and they do not completely separate the intermediate code generation and microcode generation, therefore they impose target machine and ISA dependency and do not provide backward compatibility with other target machines.

\section{abstractPIM: Three-component Code Execution Flow for PIM}
The abstractPIM flow includes two code generation components and one execution component. In the first component, \textit{intermediate representation generation}, the program is compiled into a sequence of target independent instructions based on a defined ISA. In the second component, \textit{microcode generation}, each instruction is translated into micro-operations that are supported by the target machine. The translation is performed once per instruction, and is embedded in the controller design. We adopt an existing mapping flow and modify it to support different ISAs and PIM technologies. In the third component, \textit{runtime execution}, the instructions in the compiled code are sent from the CPU to the controller, which translates them into micro-operations and sends them to the memory.

Existing logic execution methods use a set of basic logic operations to implement a logic function. They rely on a memory controller which is configured to perform these operations by applying voltages on the rows and columns of the memory array. 
In this chapter, we assume that the memory controller is configured to perform several logic operations, dubbed \textit{instructions}. Their execution sequence is determined according to a specific target machine and the PIM technology it supports. For example, if the ISA includes an AND instruction and the used technology is MAGIC NOR, 3 computation operations and 1 initialization cycle will be executed one after the other to run the AND instruction, as demonstrated in Figure~\ref{fig:controlLoad}(b), gate 1. An alternative PIM technology that consists of NAND gates will perform the same AND instruction using two NAND computations and one initialization cycle (Figure~\ref{fig:controlLoad}(c)). The instruction execution using different PIM technologies may differ in the execution time and cell usage. Our approach raises the system abstraction level and reduces the flow dependency of the specific PIM technology. It also moves one step closer towards defining a general instruction set to a memristor-based PIM architecture and designing its compiler.

The controller support of complex instructions also reduces the code size and hence the amount of code transfer between the CPU and the memory controller. However, there is a code size and execution time trade-off; the reduction in the code size may cause an execution time penalty. In a machine which supports NOR operations, the execution time, measured by the number of clock cycles in the execution sequence, is lowest when the ISA includes only NOR instructions since using basic instructions allows finer granularity. However, when using other instructions, they will be eventually executed using a NOR execution sequence. Any use of other instructions, which are, in fact, implemented using NOR micro-operations, might increase the number of NOR operations, hence the execution time. 

For example, in Figure~\ref{fig:controlLoad}, the first NOR-based implementation takes $5T_{NOR}$ clock cycles to operate, where $T_{op}$ is the number of clock cycles required for execution of an \textit{op} operation. The second implementation, however, takes a total of $T_{AND}+T_{XOR}$ clock cycles. In a machine which supports MAGIC NOR operations, the first implementation takes 10 clock cycles (2 cycles per NOR), and the second implementation takes 11 clock cycles (4 for the AND2 gate and 7 for the XOR2 gate, according to Table~\ref{table:Opcodes}). Some execution cycles are computation cycles and some are initialization cycles, as further elaborated is Section~\ref{sec:methodology}.

The instruction hierarchy in abstractPIM improves the flexibility of the compilation flow, as demonstrated in Figures~\ref{fig:controlLoad}(b)-(d). This is similar to high-level programming compared to assembly coding, which can improve flexibility at the cost of execution time penalty. While we demonstrate it using MAGIC-based logic families, the flow can be easily used for other target machines and stateful logic families. 
In our study, we choose different groups of ISAs, and different target machines that support different logic families. We demonstrate how they can be used to execute different benchmarks, and analyze the code size and execution time of the configurations.

\section{Case Study: Vector-Matrix Multiplication}
\label{sec:caseStudy}
We showcase our flow with a vector-matrix multiplication (VMM) benchmark (a 5 element vector and a 5$\times$5 matrix with 8-bit elements), which is useful in many applications, \textit{e.g.}, neural networks. The benchmark is tested over a target machine with 1024-sized memristive memory row that supports the MAGIC NOR logic family. The supported set of operations (NOT, NOR2) by the target machine is called \textit{TS0}. Other logic families are discussed in Section~\ref{sec:Results}. We first compile the benchmark for a basic case, where the ISA is also the technology set, \textit{i.e.}, \textit{TS0}. 
The selection of this ISA enabled a fair comparison between abstractPIM and existing logic execution methods, such as SIMPLER~\cite{SIMPLER}, which do not use a two-component code generation process. The used technology sets supported by the target machines we use and their instruction parameters are listed in Table~\ref{table:Opcodes}. Each instruction has three parameters: the number of inputs (\textit{I}), the number of outputs (\textit{O}) and the number of execution cycles (\textit{T}).
The first two parameters are technology independent, whereas the last parameter is technology dependent. The parameter corresponding to technology set $N$ is $T_N$. For example, the OR instruction has two inputs and a single output ($I=2$, $O=1$), and requires, when using a target machine that supports \textit{TS0}, three clock cycles for execution ($T_0=3$, two computation cycles and one initialization cycle). Using ISA=\textit{TS0} for the VMM benchmark, there are 25470 execution cycles, out of which, half are initialization cycles and half are computation cycles. Therefore, the code size is 12735 instructions.

\begin{table}[]
\centering
\scriptsize
\caption{Instruction Execution Characteristics for MAGIC Families}
\label{table:Opcodes}
\begin{threeparttable}
\begin{tabular}{|l|l|l|l|l|l|l|l|l|l|l|l|l|l|}
\hline
\rowcolor[HTML]{C0C0C0}
\multicolumn{1}{|c|}{\cellcolor[HTML]{C0C0C0}Instruction} & \multicolumn{1}{c|}{\cellcolor[HTML]{C0C0C0}I} & \multicolumn{1}{c|}{\cellcolor[HTML]{C0C0C0}O} & \multicolumn{1}{c|}{\cellcolor[HTML]{C0C0C0}$T_0$} & 
\multicolumn{1}{c|}{\cellcolor[HTML]{C0C0C0}$T_1$} & 
\multicolumn{1}{c|}{\cellcolor[HTML]{C0C0C0}$TS0$} & \multicolumn{1}{c|}{\cellcolor[HTML]{C0C0C0}$TS1$} & 
\multicolumn{1}{c|}{\cellcolor[HTML]{C0C0C0}$IS2$} & \multicolumn{1}{c|}{\cellcolor[HTML]{C0C0C0}$IS3$} \\ \hline
\multicolumn{1}{|c|}{NOT} & 
\multicolumn{1}{c|}{1} 
& \multicolumn{1}{c|}{1} & 
\multicolumn{1}{c|}{1+1} & 
\multicolumn{1}{c|}{1+1} & 
\multicolumn{1}{c|}{\cmark} & 
\multicolumn{1}{c|}{\cmark} & 
\multicolumn{1}{c|}{\cmark} & 
\multicolumn{1}{c|}{\cmark}  \\ \hline
\multicolumn{1}{|c|}{NOR2} & 
\multicolumn{1}{c|}{2} 
& \multicolumn{1}{c|}{1} & 
\multicolumn{1}{c|}{1+1} & 
\multicolumn{1}{c|}{2+1} & 
\multicolumn{1}{c|}{\cmark} & 
\multicolumn{1}{c|}{\cmark} & 
\multicolumn{1}{c|}{\cmark} & 
\multicolumn{1}{c|}{\cmark}  \\ \hline
\multicolumn{1}{|c|}{NOR3} & 
\multicolumn{1}{c|}{3} 
& \multicolumn{1}{c|}{1} & 
\multicolumn{1}{c|}{3+1} & 
\multicolumn{1}{c|}{3+1} & 
\multicolumn{1}{c|}{-} & 
\multicolumn{1}{c|}{-} & 
\multicolumn{1}{c|}{-} & 
\multicolumn{1}{c|}{\cmark}  \\ \hline
\multicolumn{1}{|c|}{NOR4} & 
\multicolumn{1}{c|}{4} 
& \multicolumn{1}{c|}{1} & 
\multicolumn{1}{c|}{5+1} & 
\multicolumn{1}{c|}{4+1} & 
\multicolumn{1}{c|}{-} & 
\multicolumn{1}{c|}{-} & 
\multicolumn{1}{c|}{-} & 
\multicolumn{1}{c|}{\cmark}  \\ \hline
\multicolumn{1}{|c|}{OR2} & 
\multicolumn{1}{c|}{2} 
& \multicolumn{1}{c|}{1} & 
\multicolumn{1}{c|}{2+1} & 
\multicolumn{1}{c|}{1+1} & 
\multicolumn{1}{c|}{-} & 
\multicolumn{1}{c|}{\cmark} & 
\multicolumn{1}{c|}{\cmark} & 
\multicolumn{1}{c|}{\cmark}  \\ \hline
\multicolumn{1}{|c|}{OR3} & 
\multicolumn{1}{c|}{3} 
& \multicolumn{1}{c|}{1} & 
\multicolumn{1}{c|}{4+1} & 
\multicolumn{1}{c|}{2+1} & 
\multicolumn{1}{c|}{-} & 
\multicolumn{1}{c|}{-} & 
\multicolumn{1}{c|}{-} & 
\multicolumn{1}{c|}{\cmark}  \\ \hline
\multicolumn{1}{|c|}{OR4} & 
\multicolumn{1}{c|}{4} 
& \multicolumn{1}{c|}{1} & 
\multicolumn{1}{c|}{6+1} & 
\multicolumn{1}{c|}{3+1} & 
\multicolumn{1}{c|}{-} & 
\multicolumn{1}{c|}{-} & 
\multicolumn{1}{c|}{-} & 
\multicolumn{1}{c|}{\cmark}  \\ \hline
\multicolumn{1}{|c|}{AND2} & 
\multicolumn{1}{c|}{2} 
& \multicolumn{1}{c|}{1} & 
\multicolumn{1}{c|}{3+1} & 
\multicolumn{1}{c|}{1+1} & 
\multicolumn{1}{c|}{-} & 
\multicolumn{1}{c|}{\cmark} & 
\multicolumn{1}{c|}{\cmark} & 
\multicolumn{1}{c|}{\cmark}  \\ \hline
\multicolumn{1}{|c|}{AND3} & 
\multicolumn{1}{c|}{3} 
& \multicolumn{1}{c|}{1} & 
\multicolumn{1}{c|}{6+1} & 
\multicolumn{1}{c|}{2+1} & 
\multicolumn{1}{c|}{-} & 
\multicolumn{1}{c|}{-} & 
\multicolumn{1}{c|}{-} & 
\multicolumn{1}{c|}{\cmark}  \\ \hline
\multicolumn{1}{|c|}{AND4} & 
\multicolumn{1}{c|}{4} 
& \multicolumn{1}{c|}{1} & 
\multicolumn{1}{c|}{9+1} & 
\multicolumn{1}{c|}{3+1} & 
\multicolumn{1}{c|}{-} & 
\multicolumn{1}{c|}{-} & 
\multicolumn{1}{c|}{-} & 
\multicolumn{1}{c|}{\cmark}  \\ \hline
\multicolumn{1}{|c|}{NAND2} & 
\multicolumn{1}{c|}{2} 
& \multicolumn{1}{c|}{1} & 
\multicolumn{1}{c|}{4+1} & 
\multicolumn{1}{c|}{2+1} & 
\multicolumn{1}{c|}{-} & 
\multicolumn{1}{c|}{-} & 
\multicolumn{1}{c|}{\cmark} & 
\multicolumn{1}{c|}{\cmark}  \\ \hline
\multicolumn{1}{|c|}{NAND3} & 
\multicolumn{1}{c|}{3} 
& \multicolumn{1}{c|}{1} & 
\multicolumn{1}{c|}{7+1} & 
\multicolumn{1}{c|}{3+1} & 
\multicolumn{1}{c|}{-} & 
\multicolumn{1}{c|}{-} & 
\multicolumn{1}{c|}{-} & 
\multicolumn{1}{c|}{\cmark}  \\ \hline
\multicolumn{1}{|c|}{NAND4} & 
\multicolumn{1}{c|}{4} 
& \multicolumn{1}{c|}{1} & 
\multicolumn{1}{c|}{10+1} & 
\multicolumn{1}{c|}{4+1} & 
\multicolumn{1}{c|}{-} & 
\multicolumn{1}{c|}{-} & 
\multicolumn{1}{c|}{-} & 
\multicolumn{1}{c|}{\cmark}  \\ \hline
\multicolumn{1}{|c|}{XOR2} & 
\multicolumn{1}{c|}{2} 
& \multicolumn{1}{c|}{1} & 
\multicolumn{1}{c|}{6+1} & 
\multicolumn{1}{c|}{5+1} & 
\multicolumn{1}{c|}{-} & 
\multicolumn{1}{c|}{-} & 
\multicolumn{1}{c|}{\cmark} & 
\multicolumn{1}{c|}{\cmark}  \\ \hline
\multicolumn{1}{|c|}{XOR3} & 
\multicolumn{1}{c|}{3} 
& \multicolumn{1}{c|}{1} & 
\multicolumn{1}{c|}{11+1} & 
\multicolumn{1}{c|}{9+1} & 
\multicolumn{1}{c|}{-} & 
\multicolumn{1}{c|}{-} & 
\multicolumn{1}{c|}{-} & 
\multicolumn{1}{c|}{\cmark}  \\ \hline
\multicolumn{1}{|c|}{XOR4} & 
\multicolumn{1}{c|}{4} 
& \multicolumn{1}{c|}{1} & 
\multicolumn{1}{c|}{16+1} & 
\multicolumn{1}{c|}{15+1} & 
\multicolumn{1}{c|}{-} & 
\multicolumn{1}{c|}{-} & 
\multicolumn{1}{c|}{-} & 
\multicolumn{1}{c|}{\cmark}  \\ \hline
\multicolumn{1}{|c|}{XNOR2} & 
\multicolumn{1}{c|}{2} 
& \multicolumn{1}{c|}{1} & 
\multicolumn{1}{c|}{5+1} & 
\multicolumn{1}{c|}{5+1} & 
\multicolumn{1}{c|}{-} & 
\multicolumn{1}{c|}{-} & 
\multicolumn{1}{c|}{\cmark} & 
\multicolumn{1}{c|}{\cmark}  \\ \hline
\multicolumn{1}{|c|}{XNOR3} & 
\multicolumn{1}{c|}{3} 
& \multicolumn{1}{c|}{1} & 
\multicolumn{1}{c|}{11+1} & 
\multicolumn{1}{c|}{6+1} & 
\multicolumn{1}{c|}{-} & 
\multicolumn{1}{c|}{-} & 
\multicolumn{1}{c|}{-} & 
\multicolumn{1}{c|}{\cmark}  \\ \hline
\multicolumn{1}{|c|}{XNOR4} & 
\multicolumn{1}{c|}{4} 
& \multicolumn{1}{c|}{1} & 
\multicolumn{1}{c|}{16+1} & 
\multicolumn{1}{c|}{8+1} & 
\multicolumn{1}{c|}{-} & 
\multicolumn{1}{c|}{-} & 
\multicolumn{1}{c|}{-} & 
\multicolumn{1}{c|}{\cmark}  \\ \hline
\multicolumn{1}{|c|}{IMPLIES} & 
\multicolumn{1}{c|}{2} 
& \multicolumn{1}{c|}{1} & 
\multicolumn{1}{c|}{2+1} & 
\multicolumn{1}{c|}{2+1} & 
\multicolumn{1}{c|}{-} & 
\multicolumn{1}{c|}{-} & 
\multicolumn{1}{c|}{\cmark} & 
\multicolumn{1}{c|}{\cmark}  \\ \hline
\multicolumn{1}{|c|}{!IMPLIES} & 
\multicolumn{1}{c|}{2} 
& \multicolumn{1}{c|}{1} & 
\multicolumn{1}{c|}{2+1} & 
\multicolumn{1}{c|}{2+1} & 
\multicolumn{1}{c|}{-} & 
\multicolumn{1}{c|}{-} & 
\multicolumn{1}{c|}{\cmark} & 
\multicolumn{1}{c|}{\cmark}  \\ \hline
\multicolumn{1}{|c|}{MUX} & 
\multicolumn{1}{c|}{3} 
& \multicolumn{1}{c|}{1} & 
\multicolumn{1}{c|}{7+1} & 
\multicolumn{1}{c|}{4+1} & 
\multicolumn{1}{c|}{-} & 
\multicolumn{1}{c|}{-} & 
\multicolumn{1}{c|}{\cmark} & 
\multicolumn{1}{c|}{\cmark}  \\ \hline
\multicolumn{1}{|c|}{HA} & 
\multicolumn{1}{c|}{2} 
& \multicolumn{1}{c|}{2} & 
\multicolumn{1}{c|}{7+1} & 
\multicolumn{1}{c|}{6+1} & 
\multicolumn{1}{c|}{-} & 
\multicolumn{1}{c|}{-} & 
\multicolumn{1}{c|}{\cmark} & 
\multicolumn{1}{c|}{\cmark}  \\ \hline
\multicolumn{1}{|c|}{HS} & 
\multicolumn{1}{c|}{2} 
& \multicolumn{1}{c|}{2} & 
\multicolumn{1}{c|}{6+1} & 
\multicolumn{1}{c|}{5+1} & 
\multicolumn{1}{c|}{-} & 
\multicolumn{1}{c|}{-} & 
\multicolumn{1}{c|}{\cmark} & 
\multicolumn{1}{c|}{\cmark}  \\ \hline
\end{tabular}%
\begin{tablenotes}[flushleft]
\setlength\tabcolsep{0pt}
\item The execution time format is $T_{c}+T_{i}$, where $T_{c}$ is the number of computation cycles and $T_{i}$ is the number of initialization cycles.
\end{tablenotes}
\end{threeparttable}
\end{table}

In attempt to reduce the code size, we used \textit{IS2}, which contains  all the functions with 1 or 2 inputs and 1 output, excluding trivial functions, \textit{e.g.}, constant '0' and identity functions\footnote{identity functions, which are in fact copy operations, can be useful in other mapping methods~\cite{SIMPLE, SAID}, but not in our row-based flow.}. The set also includes common combinational functions with more than 2 inputs or more than 1 output. Since the number of such functions is large, even for a small number of inputs, 
we chose three functions which, according to experiments we conducted, were useful in certain benchmarks: half adder [HA], multiplexer [MUX] and half subtractor [HS]. 
These instructions demonstrate the ability of our system to support blocks with more than two inputs or more than a single output. Because of the the circular dependency limitation of our flow, which is further elaborated in Section~\ref{sec:futureWork}, some useful instructions, \textit{e.g.}, 4-bit adder, could not be used. Using \textit{IS2}, code size is reduced by 52\%, but execution time is increased by 16\%.


To demonstrate the benefit of a larger number of instruction inputs and reduce the execution time, \textit{IS3} was defined. It contains the \textit{IS2} instructions, and the 2-input and single output symmetric functions from \textit{IS2} extended to 3 and 4 inputs. 
Using \textit{IS3}, lower execution time and code size, as compared to \textit{IS2}, are achieved. The execution time is increased by only 8\%, and the code size is reduced by 57\%, as compared to \textit{TS0}.

\section{abstractPIM Flow and Methodology}
\label{sec:methodology}
\begin{figure*}[!t]
    \centering
    \includegraphics[width=1\columnwidth,clip]{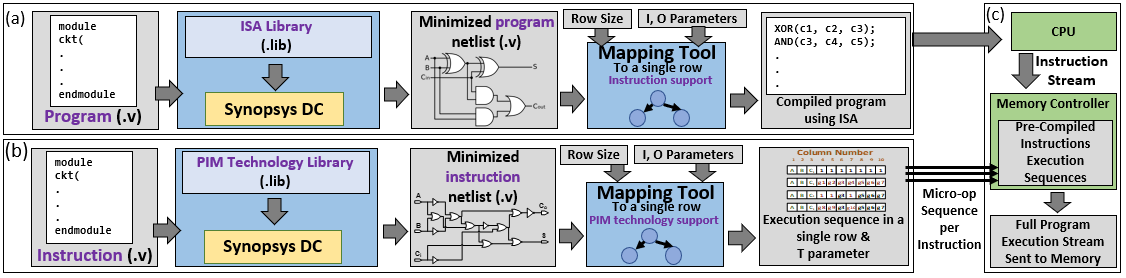}
    \caption{abstractPIM general flow is composed of three components, two components are for code generation (differences between them are marked with purple.), and the last component is for execution. 
    (a) Intermediate representation generation. (b) Microcode generation. (c) Runtime execution.}
    \label{fig:SIMPLER}
\end{figure*}

The flow of abstractPIM is composed of three components, as shown in Figure~\ref{fig:SIMPLER}. In the first component, the \textit{intermediate representation generation}, the input is a Verilog program. The program is synthesized using the Synopsys DC synthesis tool~\cite{SynopsysDC}, where the synthesis standard cell library includes the ISA in .lib format. The Synopsys DC synthesis tool was chosen since 
it supports multi-output cell synthesis. Furthermore, whereas other tools, such ABC~\cite{ABC}, support only structural Verilog, Synopsys DC supports behavioral SystemVerilog as well, therefore eases the burden of programming.
Then, a compiled program is generated using a modified and extended version of the SIMPLER mapping tool~\cite{SIMPLER}.
This tool builds a directed acyclic graph (DAG). In its original form, every node represents a NOR gate in the netlist, since SIMPLER was designed specifically for the MAGIC NOR family~\cite{MAGIC}. In the modified mapping tool, each node represents a wider variety of instructions based on the ISA. Using the DAG, the inputs and outputs of the instructions are mapped to row cells in the memristive array, and a compiled program is generated.
The \textit{I} and \textit{O} parameters are used to build the DAG and are technology-independent. The \textit{T} parameters (see Table~\ref{table:Opcodes}), which are technology-dependent and determined in the second component, are not used for compilation. Therefore, a complete separation between the code generation components and backward compatibility with other target machines is achieved.

The second component of the abstractPIM flow is \textit{microcode generation}. For each instruction, a microcode is generated by synthesizing the instruction to a micro-operation netlist and then to an execution sequence, which includes mapping to the memristive crossbar array and intermediate computation cell allocation based on specific PIM technology.
The second component input is the instruction implemented in Verilog. The instruction is synthesized using the Synopsys DC synthesis tool for a specific PIM technology, described in the synthesis standard cell library. In this chapter, we demonstrate the flow with the MAGIC~\cite{MAGIC} family, and therefore we extended the SIMPLER~\cite{SIMPLER} mapping tool to support different MAGIC operations instead of only MAGIC NOR. The execution times, listed in Table~\ref{table:Opcodes}, were calculated using this flow. The second component of abstractPIM can be replaced by handcrafted execution sequences or other mapping tools, depending on the PIM technology in use, which may produce even faster execution sequences. One such example is discussed in Section~\ref{subsec_felix}.

The general SIMPLER flow was adopted in our system for the two first components. Several modifications have been made to support different features in our tool:
\begin{enumerate}
\item \textbf{Modifications to the synthesis tool and library}. As stated above, the ABC synthesis tool~\cite{ABC} is replaced with Synopsys Design Compiler~\cite{SynopsysDC} to support synthesis with multi-output cells. The cell library format was changed to the Liberty library format, which is supported by the new synthesis tool.
\item \textbf{Modifications to the mapping tool}. While in SIMPLER each node represents a NOT or NOR operation, in abstractPIM, each node can represent a wider variety of instructions (in the first component) or micro-instructions (in the second component). Other minor changes to the SIMPLER algorithm were also performed, \textit{e.g.}, determining the \textit{FO} value of each node to include all the connected gates of each gate output and traversing the DAG accordingly, setting the \textit{CU} values according to the number of outputs of the logic gates, and parsing the new synthesis tool output.
\end{enumerate}

In the third component, \textit{runtime execution}, the two components outputs are used for full program execution. Instructions are sent from the CPU to the controller, and micro-operations are sent from the controller to the memory.

The SIMPLER mapping tool~\cite{SIMPLER} traces the number of available cells, and when they are all occupied, adds a cycle which initializes several unused cells in parallel. However, not all stateful logic families use initialization, therefore initialization cycles should not be part of the first component of the flow so we remove them. In the second component, since the flow is demonstrated using the MAGIC~\cite{MAGIC} family, we perform initialization. As opposed to SIMPLER, the second component is not aware of the full program and instruction dependencies, therefore optimized parallel initialization cannot be performed. Instead, output and intermediate computation cell initialization is performed at the first cycle of each instruction execution (if needed, additional initialization cycles can be added to the instruction execution sequence). Overall, the component separation enables flexibility and backward compatibility at the cost of execution time penalty.




In both code generation components, each standard library cell includes several parameters,  \textit{e.g.}, propagation delays and area. 
Since existing commercial synthesis tools are CMOS-oriented, we set these parameters differently and according to our memristor synthesis flow. Propagation delays, which are relevant for propagating signals in CMOS logic, are irrelevant in the context of memristor logic, where the execution time of each logic operation is a single clock cycle, and are set to 0. The area parameter is set equal for all the library cells, thus the synthesis does not prefer any particular cell, and minimizes the number of cells in the netlist,
\textit{i.e.}, minimizes the code size.

After developing abstractPIM and composing the ISAs, the code size and execution time were explored. We show the two metrics separately, due to the absence of a natural metric that combines both of them\footnote{Weighted product of code-size and execution-time was found misleading.}. It is assumed that the clock cycle time was the same for all the technology sets, so that the execution time can be measured in clock cycle units. 

We used the EPFL benchmark suite~\cite{EPFL}. These benchmarks are native combinational circuits designed to challenge modern logic optimization tools. 
The benchmark suite is divided into arithmetic and random/control parts. Each benchmark was tested with different technology sets and ISAs, listed in Table~\ref{table:Opcodes}, within a 512-sized row. One benchmark, \textit{max}, could not be mapped to a 512-sized row and was therefore tested with a 1024-sized row. 


\section{Results}
\label{sec:Results}

In this section we evaluate the abstractPIM code size reduction capabilities and execution time penalty, and discuss its abstraction, flexibility and backward compatibility advantages. 

\begin{figure}[!t]
    \setlength{\abovecaptionskip}{0pt}
    \centering
    \includegraphics[width=0.7\columnwidth,clip]{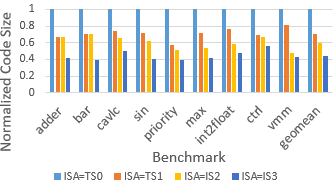}
    \caption{Normalized code size with respect to TS0 for different ISAs.
}
    \label{fig:NumOfOpcodes}
\end{figure}

The abstraction achieved by our flow using different ISAs enables backward compatibility, and 
reduces the code size as compared to an implementation based on a specific PIM technology. In the absence of a metric that measures the abstraction level achieved by our flow, we use the code size as a metric of abstraction.
Figure~\ref{fig:NumOfOpcodes} shows the code size needed for the execution of each benchmark using different ISAs: \textit{TS0}, \textit{TS1} (used as ISAs and not as technology sets), \textit{IS2} and \textit{IS3}. The code size is determined only by the ISA, and is independent of any target machine. Since the chosen sets are subsets of each other, \textit{i.e.}, $TS0 \subset TS1 \subset IS2 \subset IS3$, then $CS_{TS0}>CS_{TS1}>CS_{IS2}>CS_{IS3}$, where $CS_{set}$ is the code size of \textit{set}. Using \textit{TS1}, \textit{IS2} and \textit{IS3} reduced the code size by 30\%, 40\% and 56\% compared to \textit{TS0}, respectively.

For execution time evaluation, we compiled the benchmarks with the different ISAs and for the different target machines to demonstrate the flexibility and PIM technology independence achieved by our flow. We used two “native” configurations: \textit{TS0/TS0}, \textit{TS1/TS1}, and four “abstract” configurations: \textit{TS0/IS2}, \textit{TS0/IS3}, \textit{TS1/IS2}, and \textit{TS1/IS3}, where the notation is target-machine/ISA. We also compare the results to a single-component target-specific flow, SIMPLER~\cite{SIMPLER}.

\begin{figure}[!t]
    \setlength{\abovecaptionskip}{0pt}
    \centering
    \includegraphics[width=1\columnwidth,clip]{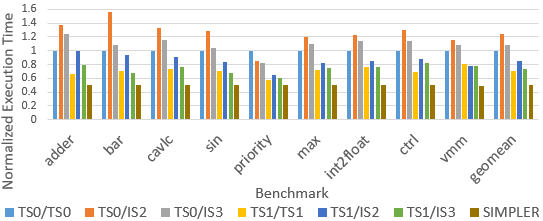}
    \caption{Normalized execution time with respect to \textit{TS0/TS0} for the different target machines and ISAs.
}
    \label{fig:ExecutionTime}
\end{figure}

The results are shown in Figure~\ref{fig:ExecutionTime}. When comparing \textit{TS0/TS0} with SIMPLER, the execution time is approximately doubled, since in our flow, every NOR or NOT operation takes an additional cycle for initialization. In SIMPLER, which operates at full program context and not at single instruction context, multiple initialization cycles can be combined and therefore the number of initialization cycles is negligible.

When comparing target machines that use native configurations (\textit{TS0/TS0} vs. \textit{TS1/TS1}) we observe that the target machine which is more capable (\textit{TS1}) runs faster (30\%). When comparing target machines that use the same abstract configuration (\textit{TS0/IS2} vs. \textit{TS1/IS2} and \textit{TS0/IS3} vs. \textit{TS1/IS3}) we also observe that the target machine which is more capable runs faster (32\% and 33\%, respectively). When comparing the execution time of a native configuration (\textit{TS0/TS0} and \textit{TS1/TS1}) with that of an abstract configuration using the same target machine, we see that the abstract configuration is slower. \textit{TS0/IS2} and \textit{TS0/IS3} are 24\% and 8\% slower than \textit{TS0/TS0}, respectively. Comparing the native \textit{TS1/TS1} configuration with the relevant abstract configurations exhibits similar results.

The above observations are quite expected. An important but less obvious benefit of abstractPIM is shown when changing a target machine. For example, when the target machine is upgraded from \textit{TS0} to \textit{TS1}, a program that has been compiled natively (\textit{TS0/TS0}) executes the same number of cycles when running on \textit{TS1} (if $TS0 \subset TS1$, otherwise even slower). However, a program that has been compiled in the first place using \textit{IS3} (\textit{IS2}) runs 27\% (16\%) faster than on the original machine – no recompilation needed. This is reflected by comparing \textit{TS1/IS3} (\textit{TS1/IS2}) vs. \textit{TS0/TS0}. 

Another observation is that among abstract ISAs, higher abstraction usually exhibits better performance, as shown by comparing \textit{TS0/IS3} vs. \textit{TS0/IS2} (13\%) and \textit{TS1/IS3} vs. \textit{TS1/IS2} (13\%). 
When comparing the results of \textit{TS0/IS2} or \textit{TS0/IS3} with \textit{TS0/TS0}, the execution time, almost always, is increased (by 24\% and 8\% for \textit{TS0/IS2} and \textit{TS0/IS3} as compared to \textit{TS0/TS0}, respectively). However, in the \textit{priority} benchmark, the execution time is decreased. On one hand, it is expected that the execution time will increase since using basic instructions allows finer granularity. On the other hand, when the number of instructions is reduced, so does the number of initialization cycles. The two opposite trends cause different benchmark behaviors. Comparison of technology \textit{TS1} and different ISAs shows the same effect. 

The flexibility and code size reduction advantages of abstractPIM come with a cost. Using MAGIC technology, in every execution cycle, one write operation is performed every clock cycle, and therefore, the number of execution cycles is also the number of write operations. The additional execution cycles per benchmark result in proportional additional energy consumption and lower effective lifetime. We believe that higher abstraction is worth the cost of these limitations. This is similar to the advantages of the abstraction achieved by a high-level programming language in comparison to low-level programming languages, \textit{e.g.}, assembly.

\section{Future Work}
\label{sec:futureWork}
In this section, we discuss future research directions that can be explored using abstractPIM, including current limitations of the abstractPIM flow and possible solutions.
\subsection{Supporting Multi-Output ISA Commands}
\label{subsec:multiOutput}
\begin{figure}[!t]
    \setlength{\abovecaptionskip}{0pt}
    \centering
    \includegraphics[width=1\columnwidth,clip]{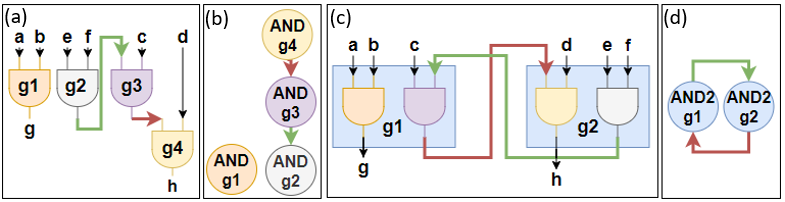}
    \caption{Compilation with multi-output instructions which creates a circular dependency.
    (a) Generated netlist using single output gate synthesis.
    (b) Generated netlist using multi-output gate synthesis.
    (c) The graph that represents netlist (a), which is a DAG and can be used for the mapping algorithm.
    (d) The graph that represents netlist (b), which includes a cyclic dependency.}
    \label{fig:bitwise}
\end{figure}

AbstractPIM supports multi-output instructions (in this work, these instructions are demonstrated as part of ISA and not supported by the target machine, since there are no multi-output MAGIC operations), but not all kinds of multi-output instructions can be used in it, since some may lead to \textit{bogus dependencies} that hinder the execution mapping. Figure~\ref{fig:bitwise} demonstrates these bogus dependencies. In the case demonstrated in Figure~\ref{fig:bitwise}, the input
is the function code: $g=ab$, $h=cdef$. In Figure~\ref{fig:bitwise}(a), the code is compiled using single-output instructions (AND2 instruction), whereas in Figure~\ref{fig:bitwise}(b), it is compiled using multi-output instructions (an instruction which computes two AND2 operations, marked in blue). Figures~\ref{fig:bitwise}(c) and~\ref{fig:bitwise}(d) show the graphs corresponding to the netlists in Figures~\ref{fig:bitwise}(a) and~\ref{fig:bitwise}(b), respectively. Whereas the graph in Figure~\ref{fig:bitwise}(c) is a DAG, the graph in Figure~\ref{fig:bitwise}(d) is not a DAG.
While there is no combinational loop in both netlists and the synthesis product is valid, a circular dependency was created between the two 2-output AND2 cells. AbstractPIM relies on the graph acyclic structure (since it uses the SIMPLER mapping algorithm), and therefore instructions which might cause cyclic dependency cannot be used. 

A sufficient condition that guarantees no such loops will be created, is to use only cells in which all the outputs depend on all the inputs, \textit{e.g.}, half adder, which implements the functions $S=a \xor b$ and $C=ab$. However, in the case of a 32-bit adder, which is a common combinational block, the first output bit $S_0$ is given by $S_0=A_0 \xor B_0 \xor C_{in}$, where $A_0$ and $B_0$ are the least significant bits of the added numbers, and $C_{in}$ is the carry in. As can be concluded from the $S_0$ calculation, it is not dependent on the other inputs and can therefore cause a cyclic graph. Future work will ensure support of any multi-output instruction, thus enabling more flexibility in planning the ISA.

\subsection{Architecture-Targeted Compilation}
When using the compilation method proposed in this chapter, the code is compiled to support different logic families, \textit{e.g.}, MAGIC NOR. This flexibility comes with a cost: the compilation does allow technology backward compatibility and execution of the code on different machines without re-compiling the code, but the compiled instruction stream is not necessarily optimized for a desired specific logic family. For example, assume a code containing a XOR logic is compiled to an ISA consisting of two instructions only: NOR and NAND. A XOR logic can be implemented, \textit{e.g.}, using either 4 NAND gates or 5 NOR gates. If the compiler is not aware of the exact target machine, it will likely compile the XOR logic into the shorter sequence consisting of 4 NAND gates. If this code is eventually run on a target machine consisting of NOR operations only, that machine implements NOR in 1 clock cycle and NAND in 4 clock cycles, so the execution will take 16 cycles (4 NAND gates) total rather than 5 cycles (5 NOR gates) total. As a result, although the code is compatible with the given target machine, it is not latency-optimized to it.

When the code is compiled for a specific stateful logic family, it can be optimized for this specific technology, \textit{e.g.}, achieving the lowest latency possible using our flow for the used technology, while maintaining backward compatibility. The optimization can be done in the first component of abstractPIM (intermediate representation), both as part of the synthesis and as part of the mapping tool. The optimization is based on technology parameters, \textit{i.e.}, the second component (microcode generation).
 
In this chapter, we discuss the optimization both as part of the synthesis and as part of the mapping tool. First, we discuss the optimization as part of the synthesis. In abstractPIM, every instruction in the ISA, which is represented by a cell in a synthesis standard library, is defined with the same area. The synthesis minimizes the area, which is equivalent in this case to minimizing the number of instructions needed for execution. However, to achieve architecture-targeted compilation, different area values can be defined for the cells in the synthesis standard library based on technology parameters. In this manner, various factors can be optimized in the synthesis. In the above NOR and NAND example, the compiler will be informed that a NAND instruction costs twice as much as a NOR instruction, and will compile the code accordingly by prioritizing the different cells in the synthesis standard library. In that sense, it is important to mention that in the case of architecture-targeted compilation, the intermediate representation and microcode generation components are no longer independent of each other. Particularly, in the case of latency optimization, the latency of each instruction, which is architecture dependent and acquired in the microcode generation component, should be embedded in the compiler. Similarly, the compiler can optimize the instruction stream considering other factors such as minimizing the number of write operations to the memristive crossbar array, prioritizing instructions with less inputs (\textit{e.g.}, prioritizing NOR2 instruction over NOR3 instruction), \textit{etc.}

Furthermore, the optimization of the aforementioned factors can be considered not only in the synthesis, but also in the mapping tool. For example, in our flow, we used SIMPLER as a mapping tool. As discussed in Section~\ref{subsec:logic_execution}, SIMPLER builds a DAG which determines an efficient gate execution order using heuristics based on different node parameters, \textit{e.g.}, \textit{CU} (cell usage). The \textit{CU} values (Equations 1 and 2) can be modified according to the instructions number of outputs and number of temporary computation cells.

As previously mentioned, the SIMPLER mapping tool can be replaced with any other mapping tool in the abstractPIM flow. Different mapping tools optimize different factors (\textit{e.g.}, latency, area, or throughput), and therefore the choice of the mapping tool depends on the architecture demands. Our flow enables to easily switch between the different mapping tools and compare them to discover the best mapping tool for a specific technology and optimization factor.
 

\subsection{High-Level Compilation}
In abstractPIM, the input to our tool is a Verilog code. This code is synthesized using a synthesis standard library that includes the ISA commands, and is eventually computed on the hardware. However, the synthesis tools used in this work are CMOS-oriented, and they are aiming for maximizing the parallelism using such a technology. These tools are not optimal for execution on a single-row memristive crossbar array with cell reuse. While abstractPIM establishes foundations for a PIM compiler, it still remains in the synthesis domain. In its current shape, abstractPIM can be used to implement instruction hierarchy in PIM architectures, but there is still work to do in order to make it a software compiler. A natural research direction is to replace the synthesis tool ("silicon compiler") with a traditional compiler ("software compiler") that compiles a high-level software code (\textit{e.g.}, Python code), into a sequence of ISA commands that can be analyzed using similar methods to those used in this chapter.  

\subsection{Supporting Input Overwriting}
\label{subsec_felix}
\begin{figure}[!t]
    \setlength{\abovecaptionskip}{0pt}
    \centering
    \includegraphics[width=0.7\columnwidth,clip]{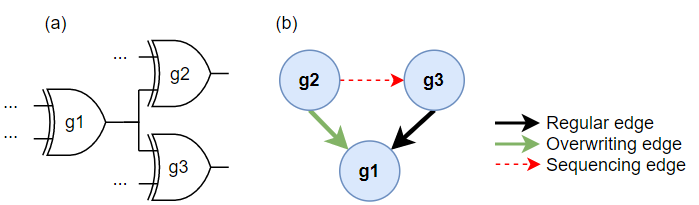}
    \caption{X-MAGIC DAG Example. (a) An example netlist of three XOR logic gates. (b) The X-MAGIC DAG corresponding to the netlist in (a). A regular edge is marked with black, an overwriting edge is marked with green and a sequencing edge is marked with red.}
    \label{fig:XMagic}
\end{figure}

As demonstrated in FELIX~\cite{felix}, single-cycle operations of different Boolean functions (and not only a NOR operation) can be implemented using MAGIC gates. The MAGIC gate inputs can be overwritten to save the utilized number of cells, to improve the effective lifetime of the system and to enhance its performance. To use such logic gates in abstractPIM, the algorithm used for the mapping should be modified to support input overwriting, as done in X-MAGIC~\cite{XMagic}. Figure~\ref{fig:XMagic} demonstrates the modifications made in the X-MAGIC DAG, which should be applied in the abstractPIM flow for overwriting support. These modifications include the definition of different edges in the DAG, each of which represents different dependency between the nodes. The first edge type is a \textit{regular edge}, which represents a non-overwriting dependency. The second edge type is an \textit{overwriting edge}, which represents a case where the output of the child node is overwritten by the parent operation.

Figure~\ref{fig:XMagic}(a) shows an example netlist that consists of three XOR logic gates, and Figure~\ref{fig:XMagic}(b) shows its corresponding X-MAGIC DAG. Gates \textit{g2} and \textit{g3} use gate \textit{g1} output, and therefore are connected to it via an edge. Assume gate \textit{g2} overwrites gate \textit{g1} output, while gate \textit{g3} does not overwrite gate \textit{g1} output. If gate \textit{g2} is executed before gate \textit{g3}, the output of gate \textit{g1} will be overwritten as part of gate \textit{g2} execution, and gate \textit{g3} will not operate properly. Therefore, regular, non-overwriting dependency (marked in black) and overwriting dependency (marked in green) are marked accordingly in the DAG. To ensure that the overwriting node is the last one executed, a sequencing dependency (marked in red) between gate \textit{g2} and gate \textit{g3} is added to the DAG.

\section{Conclusions}
This chapter presents a hierarchical compilation concept and method for logic execution within a memristive crossbar array. The proposed method provides flexibility, portability, abstraction and code size reduction. Future research directions that can enhance the abstractPIM flow, \textit{e.g.}, architecture-targeted compilation and input-overwriting support, are also presented in this chapter. 
The abstractPIM flow lays a solid foundation for a compiler for a memristor-based architecture, by enabling automatic mapping and execution of any logic function within the memory, using a defined ISA.

\section*{Acknowledgment}
This research is supported by the ERC under the European Union's Horizon 2020 Research and Innovation Programme (grant agreement no. 757259).
%
%
%
\bibliographystyle{ieeetr}

\end{document}